\documentclass[preprint,preprintnumbers,nofootinbib]{revtex4}
\usepackage[dvips]{color}
\usepackage{graphicx}

\newcommand{\diag}{{\rm diag}}
\newcommand{\ev}{{\rm eV}}

\newcommand{\gev}{{\rm GeV}}

\begin{document}

\title{Leptogenesis in a model with Friedberg-Lee symmetry}
\author{Takeshi Araki\footnote{Email:
araki@phys.nthu.edu.tw}  and 
C.~Q.~Geng\footnote{Email: geng@phys.nthu.edu.tw} }
\affiliation{Department of Physics, National Tsing Hua University,
Hsinchu, Taiwan 300.}

\begin{abstract}
We study the matter-antimatter asymmetry through the leptogenesis mechanism
in a specific model with the Friedberg-Lee (FL) symmetry.
We relate the leptogenesis with the CP violating Dirac and 
Majorana phases in the Maki-Nakagawa-Sakata leptonic mixing matrix
and illustrate the net baryon asymmetry of the universe in terms of these phases.
\end{abstract}

\maketitle

\section{Introduction}
Neutrino oscillation experiments have indicated nonzero neutrino masses and 
mixings although they are not expected in the standard model (SM).
One of the most plausible extensions of the SM to generate  neutrino 
masses is the (type-I) seesaw mechanisms in which
heavy right-handed Majorana neutrinos are introduced.
This mechanism can explain not only the small neutrino masses 
but also the baryon asymmetry of the universe (BAU) via the leptogenesis 
mechanism~\cite{LG,review}.
In the lepton sector, the leptogenesis is related to 
the CP violating Dirac and Majorana phases in the 
Maki-Nakagawa-Sakata (MNS) leptonic mixing matrix as well as some possible
high energy phases \cite{highCP}.
It is clear that there is no leptogenesis if all  phases vanish.
However, since there are many high energy phases \cite{Xing}\footnote{
In this Letter, we do not take into account the non-unitary effect of the MNS matrix.
The discussion of the leptogenesis with the effect is given in Ref. \cite{nonUni}.},
it is hard to make an explicit connection between the leptogenesis and the
phases in the MNS matrix.
In order to establish a simple relation between them, we need to reduce as many 
complex parameters as possible in the model.
In Ref. \cite{flavorLG}, a family symmetry is used to minimize the number of 
arbitrary parameters in the Yukawa sector.
Another possibility along this direction is to consider the so-called two-right-handed  
neutrino (2RHN) seesaw model \cite{3-2}, in which the number of parameters 
is less numerous than the ordinary seesaw model. 
In addition, spontaneous \cite{MChen} and dynamical \cite{YLin} 
CP violating approaches have been proposed.

In this Letter, we explore the leptogenesis mechanism in the model with the 
Friedberg-Lee (FL) symmetry \cite{FL} and directly relate it to 
the CP violating Dirac and Majorana phases in the MNS matrix. 
The FL symmetry is a translational hidden family symmetry for fermion
mass terms.
Several possible origins of the FL symmetry have been discussed in Ref. \cite{gFL}.
More detailed analyses of the symmetry have been given in Ref. \cite{FLxing}.
The FL symmetry combined with a rotational symmetry has also been
studied in Ref. \cite{twFL}.

As pointed out in Ref. \cite{jarlskog}, the introduction of the FL symmetry
to the right-handed Majorana neutrinos suggests that there exists one massless
right-handed neutrino
with the absence of  the corresponding column of the Dirac mass matrix
in the basis of the diagonal right-handed Majorana mass matrix.
As a result, the model can be regarded as the 2RHN seesaw model.
This, in fact, motivates us to examine the leptogenesis in the context of the 
FL symmetry to see if it provides us with a testable seesaw and leptogenesis model.

This Letter is organized as follows.
In Sec. II, we propose a model with the FL symmetry on both the 
right- and left-handed neutrinos.
We also examine the allowed parameters based on the present 
neutrino oscillation data.
In Sec. III, we consider leptogenesis and estimate the net baryon asymmetry of 
the universe (BAU) as a function of the Dirac and Majorana phases.
We give the conclusion in Sec. IV.

\section{Twisted Friedberg-Lee symmetric seesaw model}
\subsection{framework of model}
We start with the conventional (type-I) seesaw framework with three right-handed
Majorana neutrinos.
The relevant Lagrangian is given by
\begin{eqnarray}
 -{\cal L}=
  Y_e \bar{L}_{L} H \ell_{R}
  +Y_D \bar{L}_{L} {\tilde H} \nu_{R}
  +\frac{1}{2}M_R \overline{\nu^c}_R \nu_R+h.c.\ ,
\end{eqnarray}
where we have omitted family indeces.
We assume the diagonal charged lepton mass matrix 
and impose the twisted FL symmetry \cite{twFL} only on neutrinos
\begin{eqnarray}
 \nu_{L(R)}\rightarrow S \nu_{L(R)}+\eta \xi\ , \label{eq:tFL}
\end{eqnarray}
where $\xi$ is a non-local Grassmann parameter,
$\eta$ is a column vector of c-numbers, 
and $S$ is the permutation matrix between the second and third families,
given by
\begin{eqnarray}
 S=
 \left(\begin{array}{ccc}
 1 & 0 & 0 \\
 0 & 0 & 1 \\
 0 & 1 & 0
 \end{array}\right)\ . 
\end{eqnarray}
Here, we have adopted the specific combination $\eta \propto (-2,1,1)^T$ 
discussed in Ref. \cite{twFL} so that the resulting neutrino 
mixing is tribimaximal.
Due to the symmetry, the Majorana mass matrix takes the form
\begin{eqnarray}
M_R=
 \left(\begin{array}{ccc}
  B/2 & B/2 & B/2 \\
  B/2 & A+B & -A \\
  B/2 & -A & A+B
 \end{array}\right).\label{eq:MR}
\end{eqnarray}
In addition to the twisted FL symmetry in Eq. (\ref{eq:tFL}), we consider a $Z_2$ 
symmetry for the lepton doublet and charged singlet of the first family.
Consequently, the Dirac mass matrix is given by
\begin{eqnarray}
Y_D=
 \left(\begin{array}{ccc}
  0 & 0 & 0 \\
  0 & \alpha & -\alpha \\
  0 & -\alpha & \alpha
 \end{array}\right)\label{eq:YD}.
\end{eqnarray}
We note that without the $Z_2$ symmetry, Eq. (\ref{eq:YD}) has the same 
form as Eq. (\ref{eq:MR}).
As a result, the model cannot reproduce a realistic neutrino mass spectrum
because we focus on a scenario in which one of three light neutrinos is 
massless in this Letter.

The Majorana mass matrix in
Eq. (\ref{eq:MR}) can be diagonalized by the tribimaximal matrix \cite{TB}
\begin{eqnarray}
V_{TB}=\frac{1}{\sqrt{6}}
 \left(\begin{array}{ccc}
  2 & \sqrt{2} & 0 \\
  -1 & \sqrt{2} & -\sqrt{3} \\
  -1 & \sqrt{2} & \sqrt{3}
 \end{array}\right)\label{eq:TB},
\end{eqnarray}
so that
\begin{eqnarray}
 D_R \equiv (P V_{TB}^T) M_R (V_{TB}P) &=& \diag(M_1, M_2, M_3)\nonumber\\
  &=&\diag(0, 3/2|B|, |2A+B| ),
\end{eqnarray}
where $P=\diag(1,e^{i\phi_R /2},1)$ is a diagonal phase matrix 
of the right-handed Majorana neutrinos.
In this basis, the Dirac mass matrix in Eq. (\ref{eq:YD}) becomes
\begin{eqnarray}
Y_R\equiv Y_D V_{TB}P =\sqrt{2}\alpha
 \left(\begin{array}{ccc}
  0 & 0 & 0 \\
  0 & 0 & -1 \\
  0 & 0 & 1
 \end{array}\right)\label{eq:YR}.
\end{eqnarray}
Note that $\alpha$ can be always real by suitable redefinitions of 
the left-handed leptons.
As pointed out in Ref. \cite{jarlskog}, in this basis the right-handed neutrino of 
the first family can be regarded as a non-interacting massless neutrino.
By omitting this field, we can move to $3\times2$ dimensional Dirac mass
matrix basis and rewrite Eq. (\ref{eq:YR}) as
\begin{eqnarray}
Y_R= \sqrt{2}\alpha
 \left(\begin{array}{cc}
  0 & 0 \\
  0 & -1 \\
  0 & 1
 \end{array}\right),
\end{eqnarray}
where $D_R=\diag(M_2, M_3)$.
The mass matrix of the light neutrinos is as follows
\begin{eqnarray}
m_{\nu}=v^2 Y_R D_R^{-1} Y_R^T
=\frac{2\alpha^2 v^2}{M_3}
\left(\begin{array}{ccc}
 0 & 0 & 0 \\
 0 & 1 & -1 \\
 0 & -1 & 1
\end{array}\right).
\end{eqnarray}
This matrix can be diagonalized by the tribimaximal matrix and
has only one non-zero eigenvalue, $m_3$.
Thus, there are two interacting and one non-interacting 
massless neutrinos and  no CP violating phase in the MNS matrix.
Clearly, it is inconsistent with the experimental data of existing at least two
massive light neutrinos.

In order to obtain a realistic model, we need to 
introduce symmetry breaking terms in Eq. (\ref{eq:YD}), given by
\begin{eqnarray}
Y_D=
 \left(\begin{array}{ccc}
  0 & 0 & 0 \\
  0 & \alpha & -\alpha \\
  0 & -\alpha & \alpha
 \end{array}\right)
 +
  \left(\begin{array}{ccc}
  \frac{1}{4}(\Delta_1 + \Delta_2 + \Delta_3 + \Delta_4) & 
  \frac{1}{2}(\Delta_1 + \Delta_4) & \frac{1}{2}(\Delta_2 + \Delta_3)\\
  \frac{1}{2}(\Delta_1 + \Delta_3) & \Delta_1 & \Delta_3 \\
  \frac{1}{2}(\Delta_2 + \Delta_4) & \Delta_4 & \Delta_2
 \end{array}\right) \label{eq:YD2}.
\end{eqnarray}
Note that the breaking terms violate both the permutation symmetry in 
Eq. (\ref{eq:tFL}) and the $Z_2$ symmetry, but  preserve the 
translational symmetry
so that the first family light neutrino remains massless.
Note also that
although we could introduce breaking terms for the Majorana mass matrix as well,
we only focus on the effect from the Dirac mass matrix in the following discussions.

In the diagonal basis of the right-handed neutrinos, the Dirac mass matrix
can be still regarded as an $3\times 2$ dimensional 
matrix\footnote{This feature is ensured because the breaking terms still 
preserve the translational symmetry.}
and becomes
\begin{eqnarray}
Y_R = 
 \frac{1}{2}
  \left(\begin{array}{cc}
  \frac{\sqrt{3}}{2}(\Delta_1+\Delta_2+\Delta_3+\Delta_4) e^{i\phi_R /2} &
  -\frac{\sqrt{2}}{2}(\Delta_1-\Delta_2-\Delta_3 + \Delta_4) \\
  \sqrt{3}(\Delta_1 + \Delta_3)e^{i\phi_R /2} & 
  -2\sqrt{2}\alpha - \sqrt{2}(\Delta_1 - \Delta_3) \\
  \sqrt{3}(\Delta_2 + \Delta_4) e^{i\phi_R /2} &
  2\sqrt{2}\alpha+\sqrt{2}(\Delta_2 - \Delta_4)
 \end{array}\right).\label{eq:YR2}
\end{eqnarray}
In what follows, we consider the basis where $\alpha$ is real but $\Delta_i$
are complex and for simplicity assume $\Delta_1 \equiv \Delta \equiv |\Delta|e^{i\phi_{\Delta}}$ 
and $\Delta_2 = \Delta_3 = \Delta_4 = 0$.
The mass matrix of the light neutrinos is given by
\begin{eqnarray}
m_{\nu}= \frac{v^2}{M_3}\left[2\alpha^2
 \left(\begin{array}{ccc}
 0 & 0 & 0 \\
 0 & 1 & -1 \\
 0 & -1 & 1
 \end{array}\right)
+\frac{\alpha\Delta}{2}
 \left(\begin{array}{ccc}
 0 & 1 & -1 \\
 1 & 4 & -2 \\
 -1 & -2 & 0
 \end{array}\right)
 +\frac{\Delta^{'2}}{8}
 \left(\begin{array}{ccc}
 1 & 2 & 0 \\
 2 & 4 & 0 \\
 0 & 0 & 0
 \end{array}\right)
\right],
\label{eq:Mn}
\end{eqnarray}
where the second and third terms are responsible for the deviations from the 
tribimaximal mixing with
\begin{eqnarray}
\Delta^{'2}=\Delta^2
   \left[1+\frac{3}{2}\frac{M_3}{M_2}e^{i\phi_R} \right] ,\label{eq:Delp}
\end{eqnarray}
while $\Delta$ and $\phi_R$ generate CP violation in the MNS matrix.
Here, we define the MNS matrix as 
\begin{eqnarray}
V_{MNS}=V_{TB}\ \delta V\ \Omega=\frac{1}{\sqrt{6}}
 \left(\begin{array}{ccc}
  2 & \sqrt{2} & 0 \\
  -1 & \sqrt{2} & -\sqrt{3} \\
  -1 & \sqrt{2} & \sqrt{3}
 \end{array}\right)
 \left(\begin{array}{ccc}
 1 & 0 & 0 \\
 0 & c_{\theta} & s_{\theta} e^{-i\delta} \\
 0 & -s_{\theta} e^{i\delta} & c_{\theta}
 \end{array}\right)\Omega,\label{eq:MNS1}
\end{eqnarray}
where $s_{\theta}=\sin\theta$ ($c_{\theta}=\cos\theta$) with
\begin{eqnarray}
\tan{2\theta}=
-\frac{\sqrt{6}(\alpha\Delta+\Delta^{'2} /4)e^{i\delta}}
{(4\alpha^2+2\alpha\Delta + \Delta^{'2}/4)e^{2i\delta}-3/8\Delta^{'2} }
\equiv 
-\frac{{\cal I}e^{i\delta}}{{\cal J}e^{2i\delta}-{\cal K}}\ ,
\label{eq:theta}
\end{eqnarray}
$\delta$ is the Dirac phase which has to satisfy
\begin{eqnarray}
\delta=
-\frac{i}{2}\ln
\left[\frac{{\cal IJ^{*}+I^{*}K}}{{\cal I^{*}J+IK^{*}}}\right]
\label{eq:dirac},
\end{eqnarray}
to guarantee the right hand side of Eq. (\ref{eq:theta}) to be real, 
and $\Omega=\diag(1, e^{i\gamma /2},1)$ is a diagonal Majorana phase matrix.
The mixing angles are given by
\begin{eqnarray}
&&\sin^2\theta_{13}=\frac{1}{3}s^2_{\theta}, \label{eq:rct}\\
&&\sin^2\theta_{12} \simeq \frac{1}{3}(1-s^2_{\theta}), \label{eq:sol}\\
&&\sin^2\theta_{23} \simeq \frac{1}{2}-\frac{1}{6}s^2_{\theta}
    -\frac{\sqrt{6}}{3}s_{\theta}c_{\theta}\cos\delta. \label{eq:atm}
\end{eqnarray}
We note that our definitions of the Dirac and Majorana phases
($\delta$ and $\gamma$) are different from $\delta_{pdg}$ and $\gamma_{pdg}$ 
of the standard parametrization proposed by the
Particle Data Group \cite{pdg}. The relations between them are given by
\begin{eqnarray}
&&\cos\delta_{pdg}=
\frac{c_{12}^2 c_{23}^2 + s_{12}^2 s_{23}^2 s_{13}^2 -1/3(1-3s_{12}^2) 
     - 3/2 s_{13}^2 - \sqrt{2(1-3s_{13}^2)}\ s_{13}\cos\delta}
     {2s_{12}c_{12}s_{23}c_{23}s_{13}},\\
&&\frac{\gamma_{pdg}}{2}=\frac{\gamma}{2}+(\delta-\delta_{pdg}),
\end{eqnarray}
where $s_{ij}(c_{ij})$ means $\sin\theta_{ij}(\cos\theta_{ij})$, respectively.
The mass matrix in Eq. (\ref{eq:Mn}) is diagonalized by Eq. (\ref{eq:MNS1}), 
leading to the masses of the light neutrinos to be
\begin{eqnarray}
&&m_1=0, \\
&&m_2 = \frac{v^2}{M_3}\left|4\alpha^2 s^2_{\theta} e^{2i\delta}
+\alpha\Delta(\sqrt{6}s_{\theta}c_{\theta} e^{i\delta}
+2s^2_{\theta}e^{2i\delta})
\right.\nonumber\\ &&\hspace{4cm}\left.
+\frac{\Delta^{'2}}{4} (\sqrt{6}s_{\theta}c_{\theta} e^{i\delta}
+s^2_{\theta}e^{2i\delta}+3/2 c^2_{\theta})
\right|
 , \label{eq:m2}\\
&&m_3 = \frac{v^2}{M_3}\left|4\alpha^2 c^2_{\theta}
+\alpha\Delta(-\sqrt{6}s_{\theta}c_{\theta}e^{-i\delta}+2c^2_{\theta})
\right.\nonumber\\ &&\hspace{4cm}\left.
+\frac{\Delta^{'2}}{4} (-\sqrt{6}s_{\theta}c_{\theta} e^{-i\delta}
+3/2 s^2_{\theta}e^{-2i\delta}+c^2_{\theta})
\right|
 . \label{eq:m3}
\end{eqnarray}
The Majorana phase is given by
 \begin{eqnarray}
 \gamma&=&-\gamma_2 + \gamma_3\,,
 \label{eq:gamma}
 \end{eqnarray}
  where
\begin{eqnarray}
\sin\gamma_2 = \frac{{\rm Im} [m_2]}{|m_2|} ,\ \ 
\sin\gamma_3 = \frac{{\rm Im} [m_3]}{|m_3|}. \label{eq:majo}
\end{eqnarray}
From Eqs. (\ref{eq:dirac}), (\ref{eq:m2}), (\ref{eq:m3}), (\ref{eq:gamma}) and (\ref{eq:majo}), 
one can see that the Dirac and Majorana phases are originated from
$\phi_R$ and $\phi_\Delta$.

\subsection{low energy observables}
Our model possesses two CP violating phases: $\phi_R$ and $\phi_\Delta$,
plus four real parameters: $\alpha$, $|\Delta|$, $|2A+B|$ and $|B|$.
These six parameters can be fixed by six physical quantities.
In our numerical calculations, we use the four best-fit values of the neutrino 
oscillation data with $1\sigma$ errors \cite{osi}
\begin{eqnarray}
 &&\Delta m_{21}^2=(7.65_{-0.20}^{+0.23})\times 10^{-5}\ \ev^2,\ \ 
 |\Delta m_{31}^2|=(2.40_{-0.11}^{+0.12})\times 10^{-3}\ \ev^2,\label{eq:Dm}\\
 &&\sin^2\theta_{12}=0.304_{-0.016}^{+0.022},\ \ 
 \sin^2\theta_{23}=0.50_{-0.06}^{+0.07},\label{eq:mix}
\end{eqnarray}
as input parameters.
The remaining two quantities are the masses of the heavy neutrinos, $M_2$ and $M_3$.
As we will discuss later, in order to account for the measured value of the BAU,
they should be ${\cal O}(10^{10\sim11})\ \gev$, 
corresponding to $\alpha \sim 0.004$ and $|\Delta| \sim 0.002$.
If $M_2\gg M_3$, $\phi_R$ will be decoupled from low energy observables such as 
light neutrino masses, mixing angles and CP violating phases in the MNS matrix.
Hence, the leptogenesis ends up depending on phases 
which cannot be observed by low energy experiments.
On the other hand, in order to fit Eq. (\ref{eq:Dm}), $M_3$ cannot be 
much larger than $M_2$ and their ratio is constrained by $M_3/M_2 \leq 6$.
To illustrate our results, we take $M_3=8.0\times 10^{10}\ \gev$ and 
$M_3/M_2 = 5$.

\begin{figure}[htb]
\includegraphics*[width=0.6\textwidth]{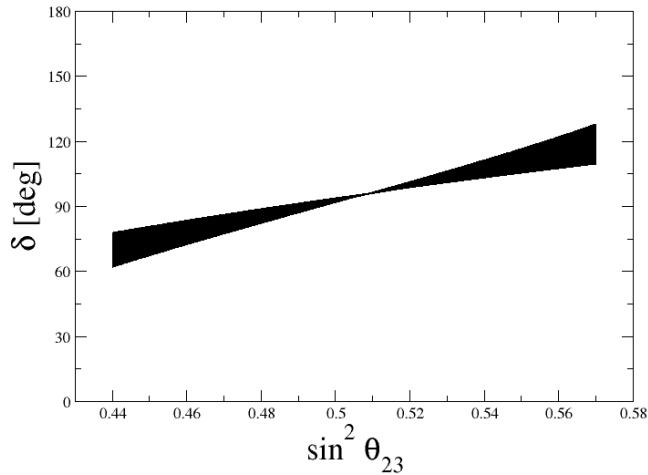}
\caption{\label{fig:a-d}\footnotesize
The Dirac phase $\delta$ as a function of $\sin^2\theta_{23}$
with $M_3=8.0\times 10^{10}\ \gev$ and $M_3/M_2 = 5$.
}
\end{figure}

From Eqs. (\ref{eq:rct}) and (\ref{eq:atm}), one can see that
the Dirac phase $\delta$ can be described by $\sin\theta_{23}$ and 
$\sin\theta_{13}$.
Furthermore, from Eqs. (\ref{eq:sol}) and (\ref{eq:mix}), one obtains that 
\begin{eqnarray}
0.0073<\sin^2\theta_{13}\,.
\label{eq:theta13}
\end{eqnarray}
By using the bound in Eq. (\ref{eq:theta13}) and $1\sigma$ values of $\sin^2\theta_{23}$, we  estimate the range of $\delta$ to be
\begin{eqnarray}
62^\circ<\delta<128^\circ\,.
\label{eq:delta}
\end{eqnarray}
In Fig. \ref{fig:a-d}, we show the numerical result of $\delta$ as a function 
of $\sin^2\theta_{23}$.
As can be seen from the figure, the result is coincident with 
Eq. (\ref{eq:delta}) very well.
We note that the above results are insensitive to the mass scale of the right-handed Majorana 
neutrinos.
\begin{figure}[htb]
\includegraphics*[width=0.6\textwidth]{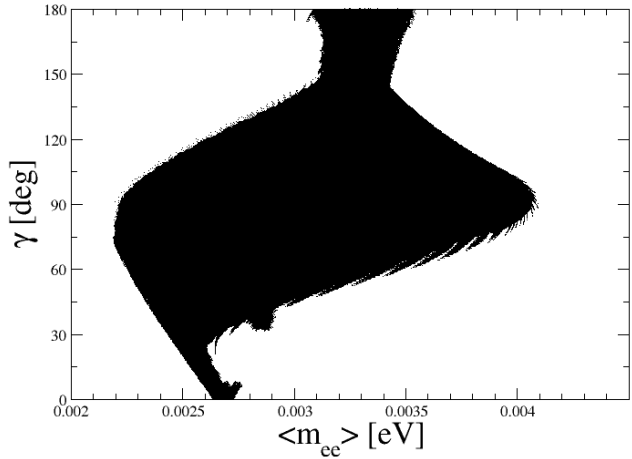}
\caption{\label{fig:0nbb}\footnotesize
$\gamma$ as a function of the effective mass $<m_{ee}>$
with $M_3=8.0\times 10^{10}\ \gev$ and $M_3/M_2 = 5$.
}
\end{figure}
In contrast, $\gamma$ has a wide allowed range and it
could have an impact on the neutrinoless double $\beta$ decay due to the 
effective Majorana mass 
\begin{eqnarray}
<m_{ee}>=\left| \sum_{i=1}^3 m_i (V_{MNS})_{1i}^2 \right|
\end{eqnarray}
since the Dirac phase as well as 
individual neutrino mass can be determined within some ranges.
In Fig. \ref{fig:0nbb}, we give $\gamma$ as a function of the effective mass.
Unfortunately, the  predicted values of $<m_{ee}>$ in our model are around
$(2.2-4.1)\times 10^{-3}$~eV, which are too small to be detected 
in the current and upcoming experiments.
For instance, the order of the present sensitivity at the CUORICINO experiment is 
$10^{-1}\ \ev$, while that of the proposed CUORE detector is 
$10^{-2}\ \ev$ \cite{0nbb}.
Nevertheless, we would like to emphasize that more dedicated experiments 
in future are needed in order to determine  the Majorana phase.

Finally, we would like to briefly remark on the possibility to test 
our model.
As our model predicts the novel relation 
$\sin^2\theta_{13}\simeq 1/3-\sin^2\theta_{12}$ based on Eqs. (\ref{eq:rct})
and (\ref{eq:sol}), more precise determinations of mixing angles would 
provide us a chance to rule out or confirm the model in future.
For instance, the smaller value of $\sin^2\theta_{12}$, which is 
$\sin^2\theta_{12}=0.304^{+0.000}_{-0.016}$, results in 
$\sin^2\theta_{13}=0.0293\sim 0.0453$ which is beyond the $1\sigma$ 
range given in Ref. \cite{osi}.
On the other hand, the larger value $\sin^2\theta_{12}=0.304^{+0.022}_{-0.000}$
corresponding to $\sin^2\theta_{13}=0.0073\sim 0.0293$ is well coincident 
with \cite{osi} and a recently proposed global analysis of the neutrino 
oscillation data \cite{s13}.

\section{Leptogenesis}
As discussed in the previous section, our model results in nonzero values of 
$\delta$ and $\sin\theta_{13}$ as shown in 
Eqs. (\ref{eq:theta13}) and (\ref{eq:delta}).
This means that the CP symmetry is always violated in the lepton sector 
at $1\sigma$ level
even if there is no Majorana phase $\gamma$. 
In this section, we consider the unflavored leptogenesis mechanism\footnote{
The importance of the flavor effects is discussed in Ref. \cite{flavor}.}
via the out-of-equilibrium decays of the heavy right-handed neutrinos.
The CP violating parameter in the leptogenesis 
 due to the i-th heavy neutrino decays is written as \cite{review}
\begin{eqnarray}
\varepsilon_i =-\frac{1}{8\pi}\sum_{j\neq i}
\frac{{\rm Im}[(Y_R^{\dag} Y_R)_{ji}^2]}{(Y_R^{\dag} Y_R)_{ii}}
\ F\left(\frac{M_j^2}{M_i^2}\right), \label{eq:gCP}
\end{eqnarray}
where $i,j=2$ or $3$, $F(x)$ is given by
\begin{eqnarray}
F(x)=\sqrt{x}\left[\frac{1}{1-x}+1-(1+x)\ln\frac{1+x}{x} \right],
\end{eqnarray}
$Y_R$ is the Dirac mass matrix in the diagonal basis of the right-handed 
neutrinos and charged leptons, given in Eq. (\ref{eq:YR2}), 
with the first (second) column  referred as $Y_{R_{j2}}$ ($Y_{R_{j3}}$).
The dilution factor $\kappa_i$ is approximately given by \cite{earlyU}
\begin{eqnarray}
 \kappa_i\simeq \frac{0.3}{r_i (\ln r_i )^{0.6}}\ ,
\end{eqnarray}
where
\begin{eqnarray}
r_i =
\frac{\Gamma_i}{H|_{T=M_i}} = \frac{M_{pl}}{1.66\sqrt{g_*}M_i^2}
\frac{(Y_R^{\dag} Y_R)_{ii}}{16\pi}M_i
\end{eqnarray}
with $M_{pl}=1.22\times 10^{19}\ \gev$ and $g_{*}=106.75$.
The net BAU is found to be
\begin{eqnarray}
\eta_B = \frac{n_B}{n_{\gamma}} = 7.04\frac{\omega}{\omega-1}
\frac{\kappa_2 \varepsilon_2 + \kappa_3 \varepsilon_3}{g_{*}},
\end{eqnarray}
where $\omega=28/79$.

We note that, in general, the CP asymmetry depends on both $\phi_R$ and 
$\phi_\Delta$ which are responsible for the Dirac and Majorana phases.
In the followings, we first examine two extreme cases of (A) $\phi_R = 0$ and 
$\phi_{\Delta}\neq 0$ and (B) $\phi_R \neq 0$ and $\phi_{\Delta} = 0$,
and then study the general case of (C) $\phi_R \neq 0$ and $\phi_{\Delta}\neq 0$.
Since $\varepsilon_2 \gg \varepsilon_3$, we will only concentrate on $\varepsilon_2$.

\subsection{$\phi_R = 0$ and $\phi_{\Delta}\neq 0$}
\begin{figure}[th]
\begin{center}
\includegraphics*[width=0.6\textwidth]{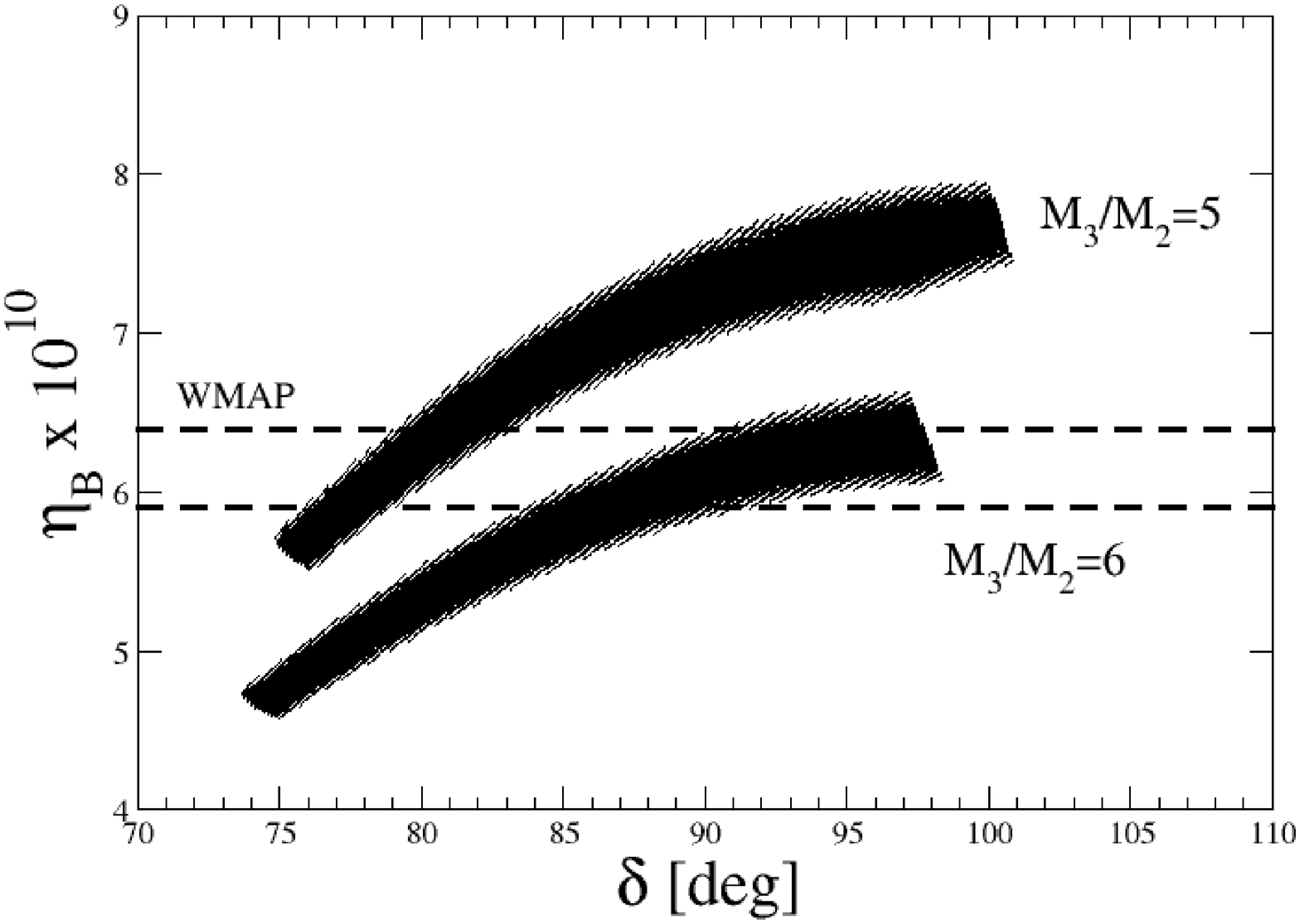}
\hspace{0.2cm}
\includegraphics*[width=0.6\textwidth]{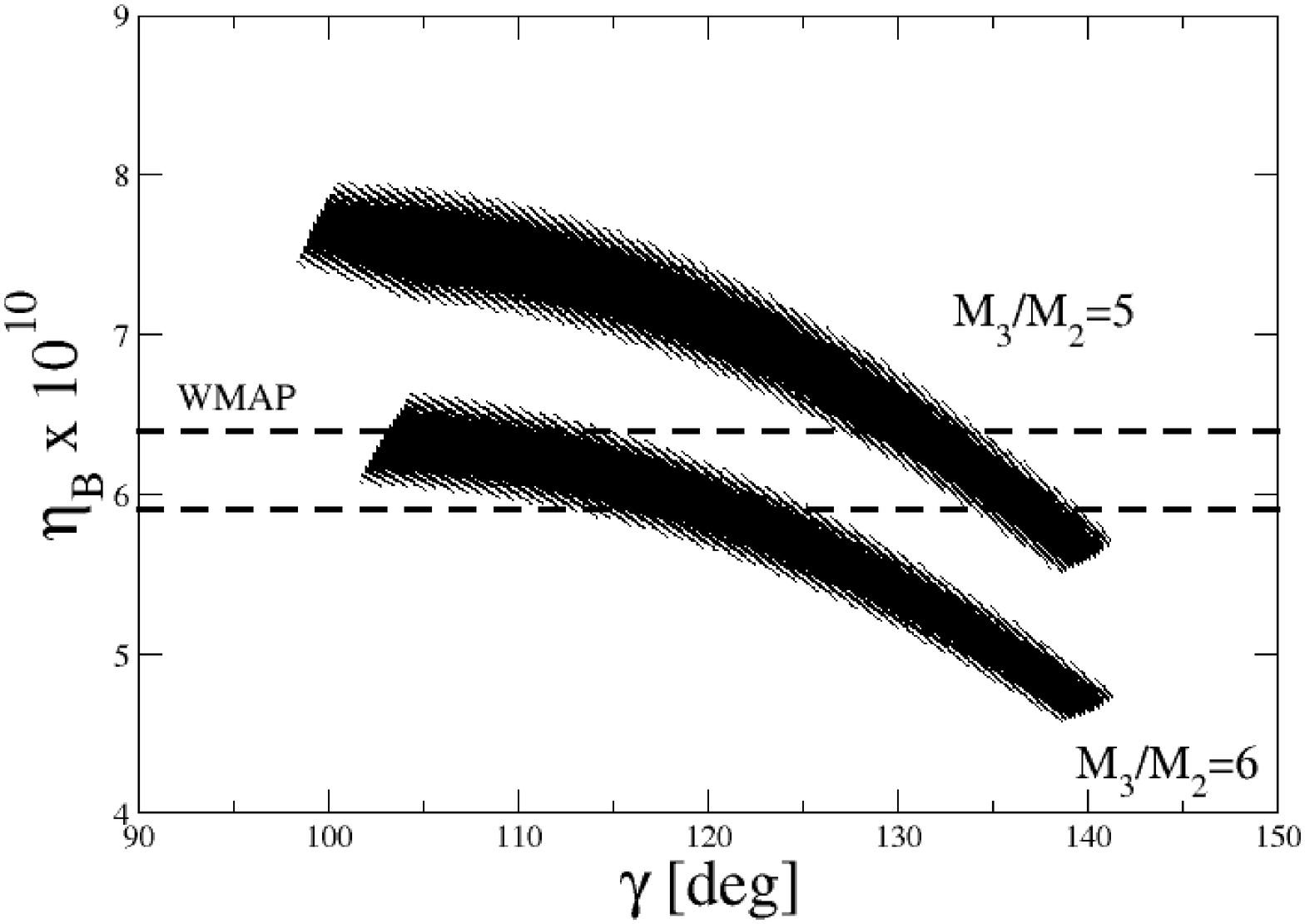}
\caption{\footnotesize
$\eta_B$ as a function of the Dirac phase $\delta$ (upper) or 
the Majorana phase $\gamma$ (lower) for $\phi_R = 0$ and $\phi_{\Delta}\neq 0$
with $M_3 = 8.0\times 10^{10}\ \gev$, where the black regions correspond to 
$M_3/M_2 =5$ and $6$, respectively, and the present $1\sigma$ WMAP bound of 
$\eta_B = (6.1_{-0.2}^{+0.2})\times 10^{10}$ 
is plotted as the dashed lines.
}\label{fig:case1}
\end{center}
\end{figure}
For $\phi_R = 0$ and $\phi_{\Delta}\neq 0$, the 
CP asymmetry
in Eq. (\ref{eq:gCP}) can be simplified as
\begin{eqnarray}
\varepsilon_2 = -\frac{1}{5\pi}\alpha\sin\phi_{\Delta}
  \left( 2\alpha\cos\phi_{\Delta}+|\Delta| \right)
  F\left(\frac{M_3^2}{M_2^2}\right).
\end{eqnarray}
In this case, from Eqs. (\ref{eq:dirac}) and (\ref{eq:majo}), one can see that
$\phi_{\Delta}$ is directly related to both Dirac and Majorana phases.
As an illustration, in Fig. \ref{fig:case1}, we estimate the net BAU in terms of
the Dirac phase $\delta$ (upper) and the Majorana phase $\gamma$ (lower) 
with $M_3 = 8.0\times 10^{10}\ \gev$, where the regions correspond to 
$M_3/M_2 =5$ and $6$, respectively, and the 
present WMAP bound of $\eta_B = (6.1_{-0.2}^{+0.2})\times 10^{10}$ 
\cite{wmap} at $1\sigma$ is plotted as the dashed lines.
 From the figure, to obtain the measured BAU at $1\sigma$, we find that 
 $\delta\sim 76^\circ-83^\circ$ ($84^\circ-98^\circ$) and 
 $\gamma\sim 127^\circ-140^\circ$ 
 ($102^\circ-125^\circ$) for $M_3/M_2 =5~(6)$. 
 The allowed ranges of the phase parameters for 
 $M_3/M_2 =5$ are smaller than those for $M_3/M_2 =6$.

\subsection{$\phi_R \neq 0$ and $\phi_{\Delta}=0$}
\begin{figure}[th]
\begin{center}
\includegraphics*[width=0.6\textwidth]{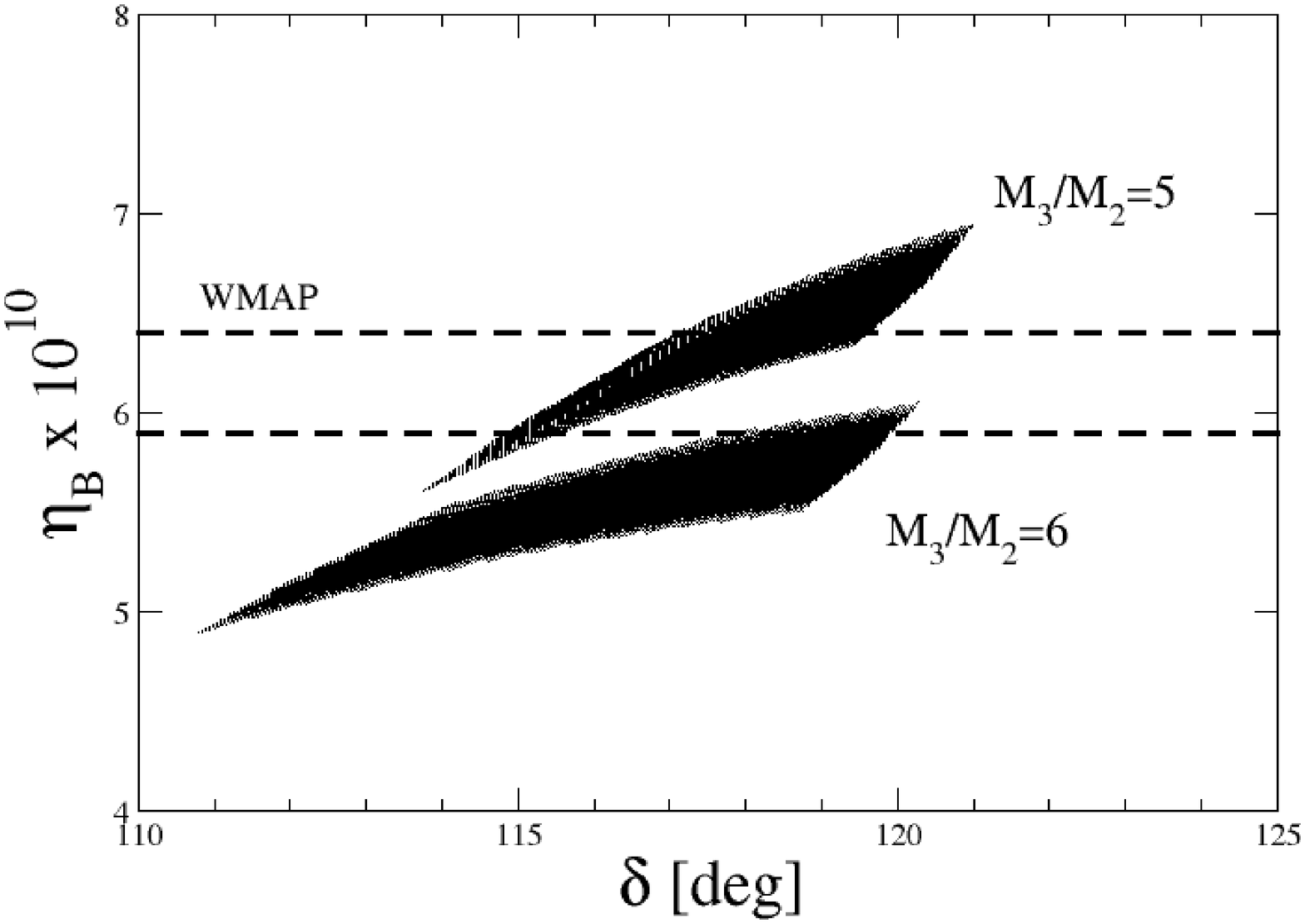}
\hspace{0.2cm}
\includegraphics*[width=0.6\textwidth]{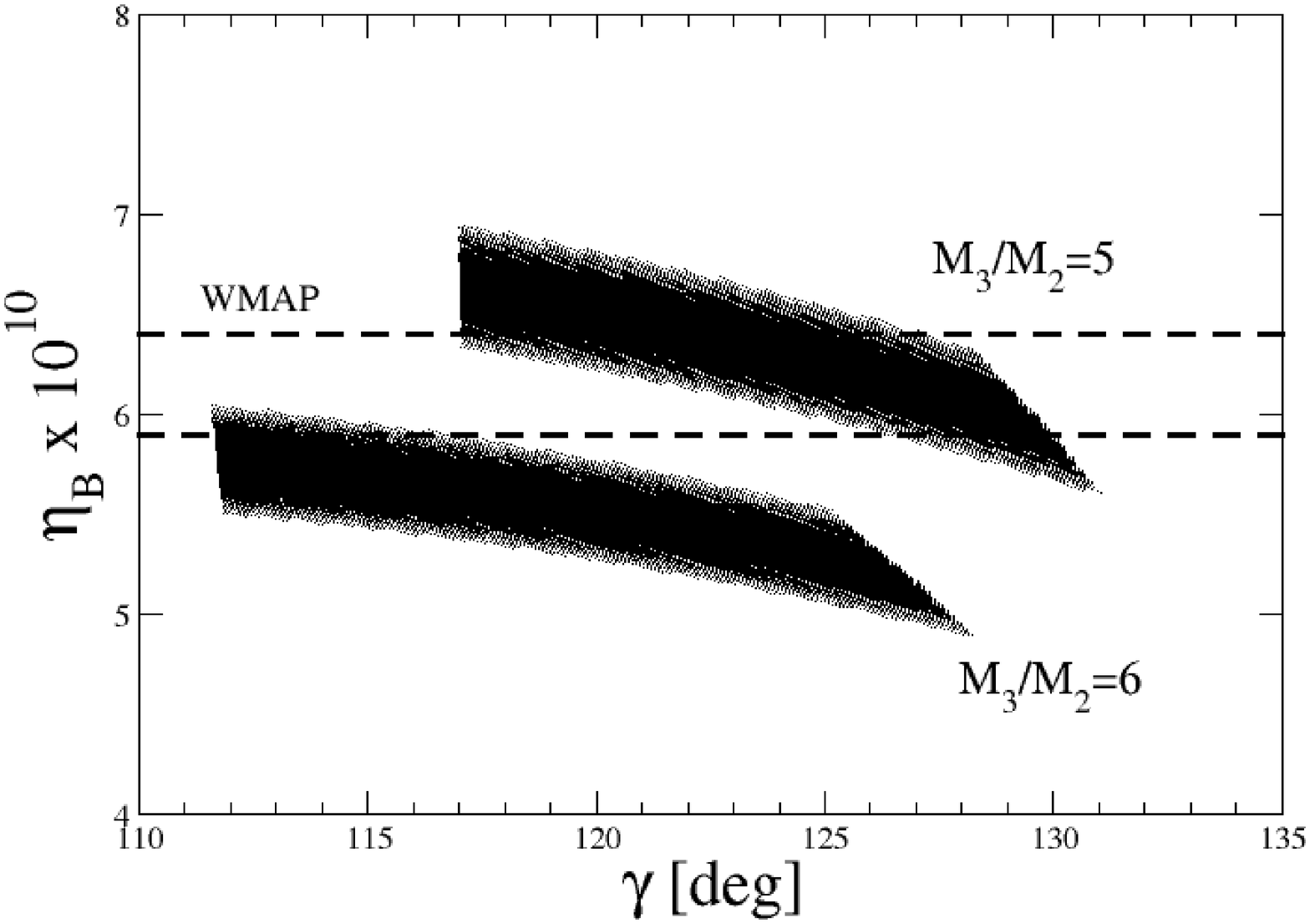}
\caption{\footnotesize
Legend is the same as Fig. \ref{fig:case1} but for 
$\phi_R \neq 0$ and $\phi_{\Delta}=0$.
}\label{fig:case2}
\end{center}
\end{figure}
Similarly, for $\phi_R \neq 0$ and $\phi_{\Delta}=0$, the CP asymmetry is given by 
\begin{eqnarray}
\varepsilon_2 = -\frac{1}{20\pi}\sin\phi_R
  \left[ \frac{|\Delta|^2}{16} + (|\Delta| + 2\alpha)^2 \right]
  F\left(\frac{M_3^2}{M_2^2}\right),
\end{eqnarray}
where the phase $\phi_{R}$ is responsible for both Dirac and Majorana phases.
In Fig. \ref{fig:case2}, we display $\eta_B$ as a function of $\delta$ 
(upper) or $\gamma$ (lower) similar to Fig. \ref{fig:case1}.
 As seen from the figure,  at $1\sigma$ level,
  almost all phase parameters in the plane for $M_3/M_2 =6$ are
 ruled out by the WMAP data of the BAU, whereas 
for $M_3/M_2 =5$, $\delta$ and $\gamma$ are found to be around
$115^\circ - 120^\circ$ and $117^\circ - 130^\circ$, respectively.

\subsection{$\phi_R \neq 0$ and $\phi_{\Delta}\neq 0$}
\begin{figure}[htb]
\includegraphics*[width=0.4\textwidth]{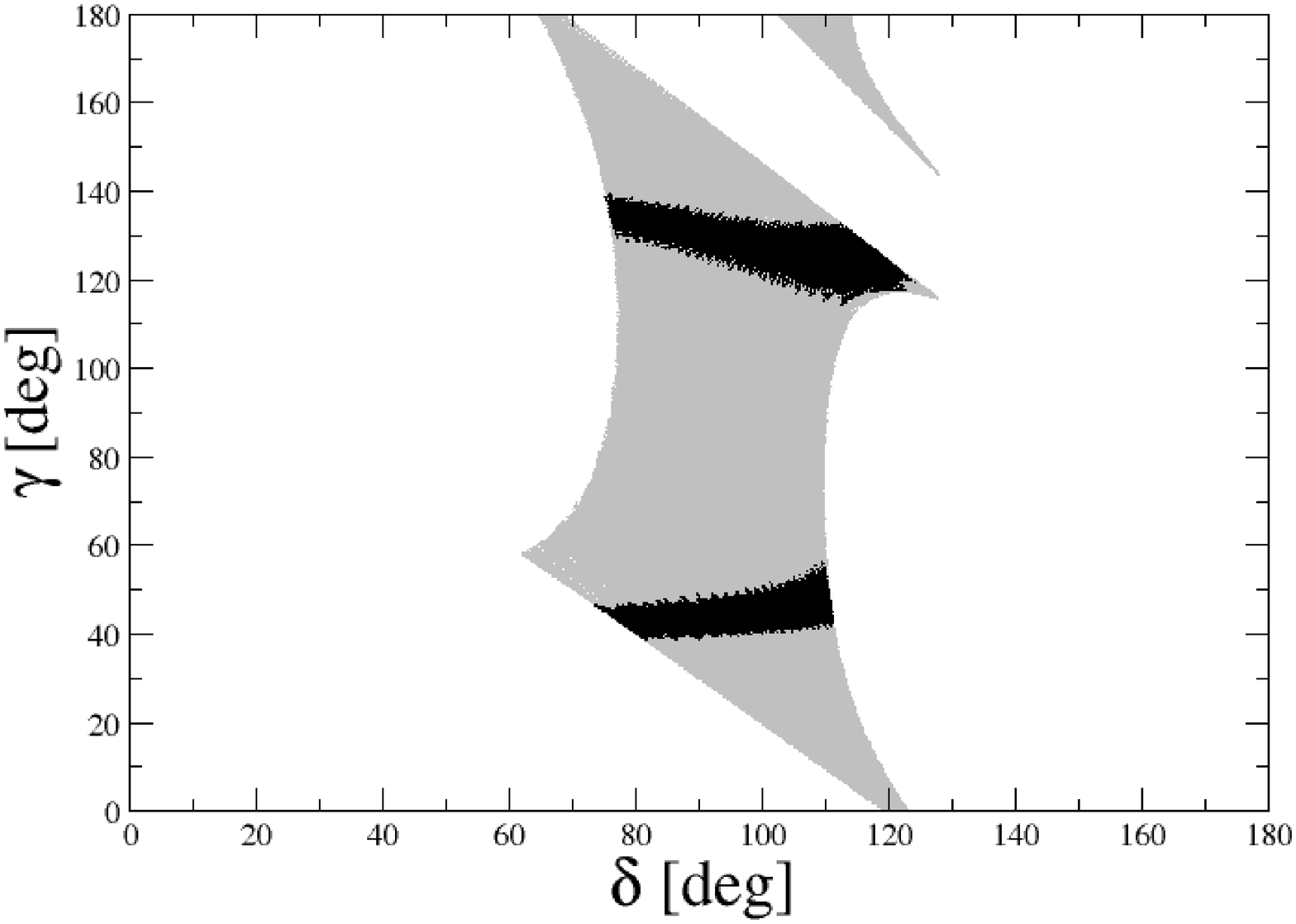}
\hspace{0.2cm}
\includegraphics*[width=0.4\textwidth]{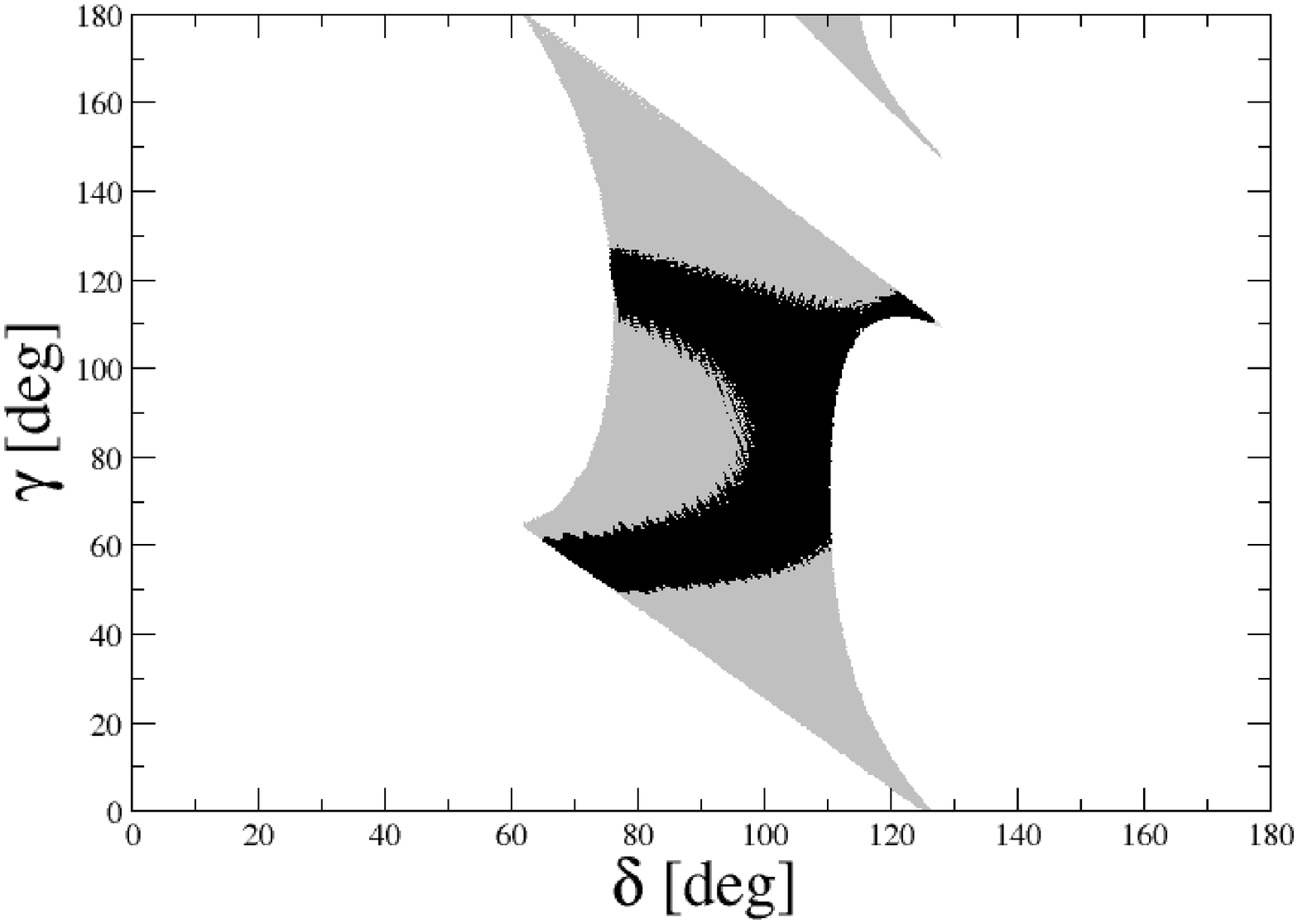}
\caption{
\footnotesize
Allowed regions in $\delta-\gamma$ plane
with $M_3=8.0\times 10^{10}\ \gev$
and $M_3/M_2 = 5$ (left) and $6$ (right), 
where the gray and black regions correspond to
those fitted by the neutrino oscillation  and
WMAP data at $1\sigma$, respectively.
 }\label{fig:d-m}
\end{figure}
For the general case of $\phi_R \neq 0$ and $\phi_{\Delta}\neq 0$, instead of 
showing a much more complex analytic formula of $\varepsilon_2$, 
we only give the numerical results.
In Fig. \ref{fig:d-m}, 
we show the allowed regions in $\delta-\gamma$ plane
with $M_3=8.0\times 10^{10}\ \gev$
and $M_3/M_2 = 5$ (left) and $6$ (right), 
where the gray and black regions represent to
those fitted by the neutrino oscillation  and 
WMAP data, respectively.
In contrast with the previous cases of (A) and (B), two narrow ranges of 
$\gamma$ and a wide range of  $\delta$ are allowed for $M_3/M_2 =5$,
while those for $M_3/M_2 =6$ are continuous and broad.

\section{conclusion}
We have studied the BAU through the leptogenesis mechanism
in the model with the FL symmetry.
We have tried to make a connection between the leptogenesis and
 the CP violating Dirac and 
Majorana phases in the MNS  matrix.
In particular, 
we have demonstrated that there exists 
a wide range of these phases to achieve the measure BAU, allowed by
the neutrino data.
\\

\noindent {\bf Acknowledgments}

This work is supported in part by 
the National Science Council of ROC under Grant No:
 NSC-95-2112-M-007-059-MY3 and by the Boost Program of NTHU.


\begin{thebibliography}{3}
\bibitem{LG}
M. Fukugita and T. Yanagida, Phys. Lett. B {\bf 174}, 45 (1986).

\bibitem{review}
For recent reviews on the leptogenesis, see 
M. C. Chen, TASI 2006 Lectures on Leptogenesis, hep-ph/0703087;
S. Davidson, E. Nardi and Y. Nir, Phys. Rept. {\bf 466}, 105 (2008).

\bibitem{highCP}
G. C. Branco, T. Morozumi, B. M. Nobre and M. N. Rebelo, 
Nucl. Phys. B {\bf 617}, 475 (2001). 

\bibitem{Xing}
For a recent review, see 
Z. Z. Xing, arXiv:0902.2469 [hep-ph].

\bibitem{nonUni}
W. Rodejohann, arXiv:0903.4590 [hep-ph].

\bibitem{flavorLG}
T. Araki, J. Kubo and E. A. Paschos, Eur. Phys. J. C {\bf 45}, 465 (2006);
R. N. Mohapatra and H. B. Yu, Phys. Lett. B {\bf 644}, 346 (2007).

\bibitem{3-2}
P. H. Frampton, S. L. Glashow and T. Yanagida, Phys. Lett. B {\bf 548}, 119 (2002);
T. Endoh, S. Kaneko, S. K. Kang, T. Morozumi and M. Tanimoto, 
Phys. Rev. Lett. {\bf 89}, 231601 (2002);
T. Kitabayashi, Phys. Rev. D {\bf 76}, 033002 (2007);
Prog. Theor. Phys. {\bf 120}, 443 (2008);
B. Brahmachari and N. Okada, Phys. Lett. B {\bf 660}, 508 (2008).

\bibitem{MChen}
M. C. Chen and K. T. Mahanthappa, Phys. Rev. D {\bf 71}, 035001 (2005).

\bibitem{YLin}
Y. Lin, arXiv:0903.0831 [hep-ph].

\bibitem{FL}
R. Friedgerg and T. D. Lee, High Energy Phys. Nucl. Phys. {\bf 30}, 591 (2006);
Annal Phys. {\bf 323}, 1087 (2008).

\bibitem{gFL}
C. S. Huang, T. J. Li, W. Liao and S. H. Zhu, Phys. Rev. D {\bf 78}, 013005 (2008). 

\bibitem{FLxing}
Z. Z. Xing, H, Zhang and S. Zhou, Phys. Lett. B {\bf 641}, 189 (2006);
S. Luo and Z. Z. Xing, Phys. Lett. B {\bf 646}, 242 (2007);
Phys. Rev. D {\bf 78}, 117301 (2008). 

\bibitem{twFL}
T. Araki and R. Takahashi, arXiv:0811.0905 [hep-ph].

\bibitem{jarlskog}
C. Jarlskog, Phys. Rev. D {\bf 77}, 073002 (2008).

\bibitem{TB}
P. F. Harrison, D. H. Perkins and W. G. Scott, Phys. Lett. B {\bf 530}, 167 (2002);
P. F. Harrison and W. G. Scott, Phys. Lett. B {\bf 535}, 163 (2002).

\bibitem{pdg}
C. Amsler {\it et al}. [Particle Data Group], Phys. Lett. B {\bf 667}, 1 (2008).  

\bibitem{osi}
T. Schwetz, M. Tortola and J. W. F. Valle, New J. Phys. {\bf 10}, 113011 (2008).

\bibitem{s13}
 G. L. Fogli, E. Lisi, A. Marrone, A. Palazzo and A. M. Rotunno,
 Phys. Rev. Lett. {\bf 101}, 141801 (2008); arXiv:0905.3549 [hep-ph].

\bibitem{0nbb}
C. Arnaboldi {\it et al}. [CUORICINO Collaboration], 
Phys. Rev. C {\bf 78}, 035502 (2008).

\bibitem{flavor}
A. Abada, {\it et al}.,
JCAP {\bf 0604}, 004 (2006);
E. Nardi, Y. Nir, E. Roulet and J. Racker, JHEP {\bf 0601}, 164 (2006);
A. Abada, {\it et al}., JHEP {\bf 0609}, 010 (2006).

\bibitem{earlyU}
E. W. Kolb and M. S. Turner, "{\it The early universe}", Addison-Wesley (1990) (Frontiers in physics, 69).

\bibitem{wmap}
J. Dunkley {\it et al}. [WMAP Collaboration], Astrophys. J. Suppl. {\bf 180}, 306 (2009).

\end{thebibliography}
\end{document}